\documentclass{aa}  

\usepackage{graphicx}
\usepackage{txfonts}
\usepackage{lscape}
\usepackage{array}

\newcommand\clearrow{\global\let\rowmac\relax}
\clearrow
\usepackage{hyperref}
\hypersetup{
    colorlinks=true,
    linkcolor=blue,
    citecolor=blue
}
\begin{document} 

    \title{Magnetic reconnection as a mechanism to produce multiple proton populations and beams locally in the solar wind}
    \titlerunning{Magnetic reconnection as a mechanism to produce multiple proton populations and beams}

   \author{B. Lavraud \inst{1,2} \and R. Kieokaew \inst{2} \and N. Fargette \inst{2} \and P. Louarn \inst{2}
          \and A. Fedorov \inst{2}
          \and N. André \inst{2}
          \and G. Fruit \inst{2}
          \and V. Génot \inst{2}
          \and V. Réville \inst{2}
          \and A. P. Rouillard \inst{2}
          \and I. Plotnikov \inst{2}
          \and E. Penou \inst{2}
         \and A. Barthe \inst{3} 
         \and L. Prech \inst{4} 
         \and C. J. Owen \inst{5} 
         \and R. Bruno \inst{6} 
         \and F. Allegrini \inst{7,8} 
         \and M. Berthomier \inst{9} 
         \and D. Kataria \inst{5} 
         \and S. Livi \inst{7,8} 
         \and J. M. Raines \inst{10} 
         \and R. D’Amicis \inst{6} 
         \and J. P. Eastwood \inst{11} 
         \and C. Froment \inst{12} 
         \and R. Laker \inst{11} 
         \and M. Maksimovic \inst{13} 
         \and F. Marcucci \inst{6} 
         \and S. Perri \inst{14} 
         \and D. Perrone \inst{15} 
         \and T. D. Phan \inst{16} 
         \and D. Stansby \inst{5} 
         \and J. Stawarz \inst{11} 
         \and S. Toledo-Redondo \inst{17} 
         \and A. Vaivads \inst{18}
         \and D. Verscharen \inst{5, 19} 
         \and I. Zouganelis \inst{19} 
         \and V. Angelini \inst{11} 
         \and V. Evans \inst{11} 
         \and T. S. Horbury \inst{11} 
         \and H. O’brien \inst{11}
          }
   \institute{Laboratoire d'astrophysique de Bordeaux, Univ. Bordeaux, CNRS, Pessac, France 
%   \\
%             \email{benoit.lavraud@u-bordeaux.fr}
              \and Institut de Recherche en Astrophysique et Planétologie, CNRS, UPS, CNES, Toulouse, France \and AKKA-Toulouse, France \and 
Department of Surface and Plasma Science, Faculty of Mathematics and Physics, Charles University, Prague, Czech Republic \and 
Mullard Space Science Laboratory, University College London, Holmbury St. Mary, Dorking, Surrey, UK \and 
 INAF-Istituto di Astrofisica e Planetologia Spaziali, Via Fosso del Cavaliere 100, 00133 Roma, Italy \and 
 Southwest Research Institute, San Antonio, USA \and 
 Department of Physics and Astronomy, University of Texas at San Antonio, San Antonio, Texas, USA \and
 Laboratoire de Physique des Plasmas, Ecole Polytechnique, Palaiseau, France \and 
 Department of Climate and Space Sciences and Engineering, The University of Michigan, Ann Arbor, USA \and 
Space and Atmospheric Physics, The Blackett Laboratory, Imperial College London, London, UK \and 
LPC2E, CNRS, University of Orléans, CNES, Orléans, France \and 
LESIA, Meudon, France \and 
Dipartimento di Fisica, Universita della Calabria, Italy \and 
ASI - Italian Space Agency, Italy \and 
Space Sciences Laboratory, University of California, Berkeley, USA \and 
University of Murcia, Spain \and 
KTH, Stockholm, Sweden \and 
European Space Agency (ESA), European Space Astronomy Centre (ESAC), Camino Bajo del Castillo s/n, 28692 Villanueva de la Cañada, Madrid, Spain \and 
Space Science Center, University of New Hampshire, Durham NH 03824, USA
 \\            }

   \date{Received April 16, 2021}

% \abstract{}{}{}{}{} 
% 5 {} token are mandatory
 
  \abstract
  % context heading (optional)
  % {} leave it empty if necessary  
   {Spacecraft observations early revealed frequent multiple proton populations in the solar wind. Decades of research on their origin have focused on processes such as magnetic reconnection in the low corona and wave-particle interactions in the corona and locally in the solar wind.}
  % aims heading (mandatory)
   {This study aims to highlight that multiple proton populations and beams are also produced by magnetic reconnection occurring locally in the solar wind.}
  % methods heading (mandatory)
   {We use high resolution Solar Orbiter proton velocity distribution function measurements, complemented by electron and magnetic field data, to analyze the association of multiple proton populations and beams with magnetic reconnection during a period of slow Alfvénic solar wind on 16 July 2020. }
  % results heading (mandatory)
   {At least 6 reconnecting current sheets with associated multiple proton populations and beams, including a case of magnetic reconnection at a switchback boundary, are found during this day. This represents 2\% of the measured distribution functions. We discuss how this proportion may be underestimated, and how it may depend on solar wind type and distance from the Sun.}
  % conclusions heading (optional), leave it empty if necessary 
   {Although suggesting a likely small contribution, but which remains to be quantitatively assessed, Solar Orbiter observations show that magnetic reconnection must be considered as one of the mechanisms that produce multiple proton populations and beams locally in the solar wind.}
   \keywords{solar wind – proton beams – magnetic reconnection – turbulence – wave-particle interaction}
   \maketitle
%%\watermark{DRAFT}
%\graphicspath{{./}{figures/}}
%%%%%%%%%%%%%%%%%%%%%%%%%%%%%%%%%%%%%%%%%%%%%%%%%%%%%%%%%%%%%%%%%%%%%%%%%%%%%%
%% We recommend that authors also use the natbib \citep
%% and \citet commands to identify citations.  The citations are
%% tied to the reference list via symbolic KEYs. The KEY corresponds
%% to the KEY in the \bibitem in the reference list below. 

%%%%%%%%%%%%%%%%%%%%%%%%%%%%%%%%%%%%%%%%%%%%%%%%%%%%%%%%%%%%%%%%%%%%%%%%%%%%%%
\section{Introduction} \label{sec:1_intro}
%%%%%%%%%%%%%%%%%%%%%%%%%%%%%%%%%%%%%%%%%%%%%%%%%%%%%%%%%%%%%%%%%%%%%%%%%%%%%%

The observation of solar wind proton distribution functions made of two components, a core and a beam population, was reported early in the space exploration era \citep{FELDMAN_1973}. \citet{FELDMAN_1974, 1996A&A...316..355F} argued that the proton beams, typically of lower intensity than the core, might stem from proton injections into the nascent solar wind in the low corona or at chromospheric level. They pointed to the likely role of magnetic reconnection between open field lines and closed loops in the low corona, a mechanism widely used to explain various types of coronal and solar wind observations, such as the magnetic switchbacks that got recent strong focus with Parker Solar Probe (PSP) data \citep[e.g.][]{Kasper_2019, Bale_2019} (cf. Belcher and Davies (1971) for early observations of large-scale Alfvénic structures akin to switchbacks).

A very large body of work, based on observations, theory and simulations, has investigated the origin of multiple proton populations and beams in the solar wind. Most recent studies have led to a paradigm shift whereby the proton multiple populations and beams are born out of wave-particle interactions (of various types) and turbulence \citep[e.g.][]{MONTGOMERY_1976, LIVI_1987, GARY_1991, DAUGHTON_GARY_1998, DAUGHTON_1999, TAM_1999, TU_2002, TU_2004, ARANEDA_2008, MATTEINI_2010, OSMANE_2010, PIERRARD_2010, VALENTINI_2011, VOITENKO_2015, ALTERMAN_2019}. Many studies also focused on the instabilities that result from the presence of the beam, and on their effects back on the proton distributions themselves \citep[e.g.][]{WONG_1988, GOMBEROFF_2006, MATTEINI_2010, HELLINGER_2011, HELLINGER_2013, MATTEINI_2015, CHEN_2016, WICKS_2016, VERSCHAREN_2016, SHAABAN_2020, KLEIN_2021, LOUARN_2021}. 

Yet, in parallel it was early realized that reconnection produces a mixing of particle populations from the upstream inflow regions (the boundary conditions of the reconnecting current sheet), leading among other effects to the production of proton distribution functions comprising multiple populations and beams \citep[e.g.][]{LOTTERMOZER_1998, SHAY_1998, HOSHINO_1998, PHAN_2007, DRAKE_2009, AUNAI_2011, EASTWOOD_2015, INNOCENTI_2017}. 

Apart from \citet{CHEN_2016}, who do list reconnection as a possible source for proton beams in the solar wind (they cite \citet{GOSLING_2005}), our thorough but yet non-exhaustive bibliographic search did not reveal any other work indicating magnetic reconnection as a local mechanism to produce proton beams in the solar wind. \citet{FARRUGIA_2001} and \citet{GOSLING_2005} demonstrated that magnetic reconnection occurs locally in the solar wind. \citet{GOSLING_2005} specifically showed that the transport of plasmas from the two upstream regions produces interpenetrating proton beams in the exhaust. This is a demonstration that at least a portion of multiple ion populations in the solar wind is produced locally by magnetic reconnection. Since then, among the numerous studies of magnetic reconnection in the solar wind \citep[e.g.][]{PHAN_2006, DAVIS_2006, LAVRAUD_2009, TIAN_2010, FENG_2011, XU_2011, ENZL_2014, MISTRY_2017, FENG_2017, HE_2018, EASTWOOD_2018, KHABAROVA_2021}, none have focused on the multiple proton population aspect apart from \citet{Huttunen_2008}, who observed secondary proton enhancements consistent with a beam in the vicinity of the separatrix (a fact also reported here). 

The \citet{GOSLING_2005} event was special because it occurred within a Coronal Mass Ejection (CMES) with low $\beta$ and high Alfvén speed. Because the separation between the two beams scales with the Alfvén speed (cf. section~\ref{sec:2_obs}), the high Alfvén speed allowed the interpenetrating beams to be easily distinguished, whereas for regular higher $\beta$ solar wind their identification requires higher energy and angular ion measurements, or may not be distinguishable at all. In the present paper, we report further cases showing multiple proton populations and beams in the solar wind as a result of local magnetic reconnection, both in the exhaust and in the boundary layers on the outside, thanks to the new high-resolution ion measurements onboard Solar Orbiter. We purport that magnetic reconnection is a mechanism that must be considered when investigating the origin of multiple proton populations and beams in the solar wind.

%%%%%%%%%%%%%%%%%%%%%%%%%%%%%%%%%%%%%%%%%%%%%%%%%%%%%%%%%%%%%%%%%%%%%%%%%%%%%%
\section{Observations} \label{sec:2_obs}
%%%%%%%%%%%%%%%%%%%%%%%%%%%%%%%%%%%%%%%%%%%%%%%%%%%%%%%%%%%%%%%%%%%%%%%%%%%%%%

\begin{figure*}[ht]
\centering
\includegraphics[width=1\textwidth]{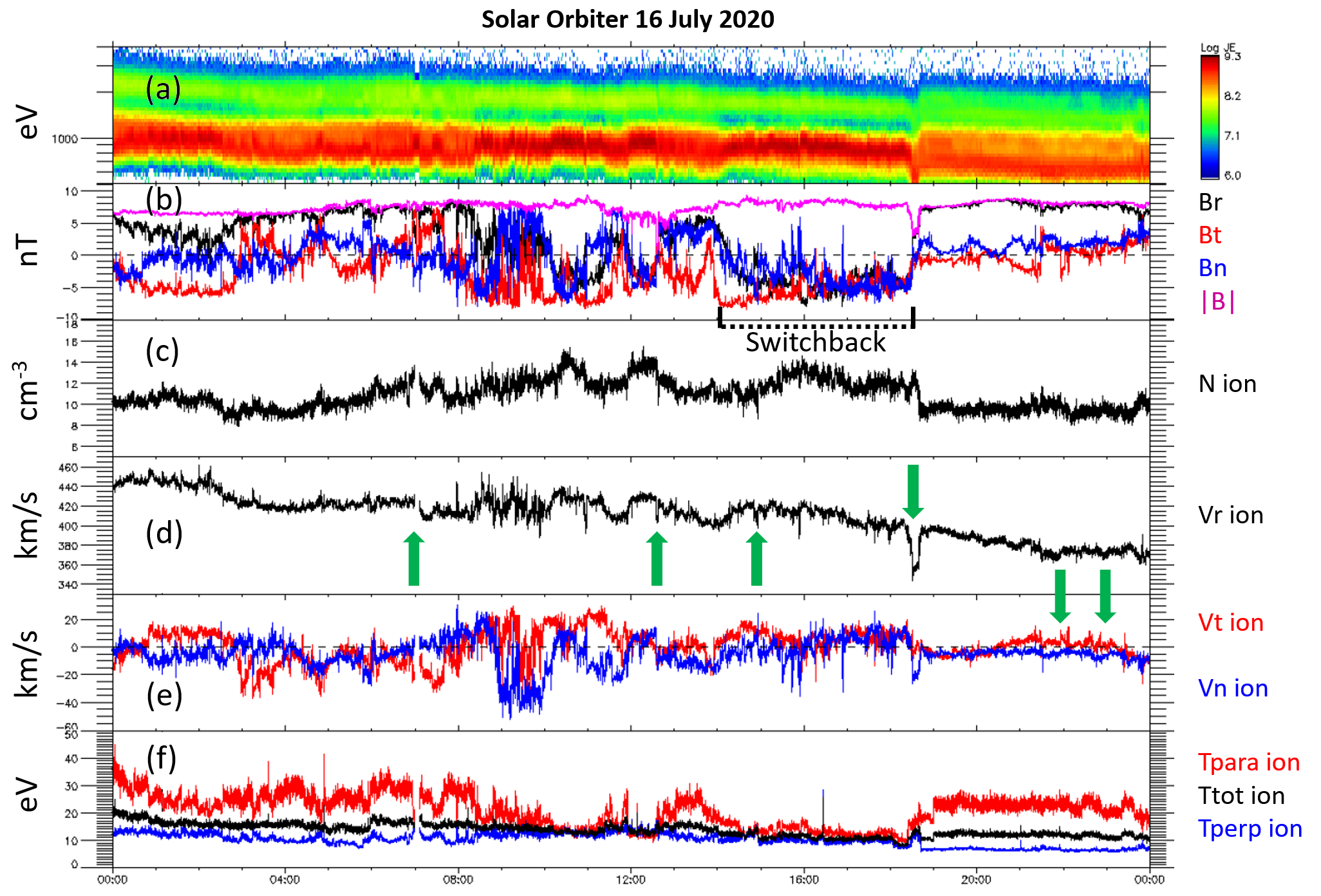}
\caption{Solar Orbiter ion and magnetic field observations on 16 July 2020. (a) Energy time spectrogram for all ions. (b) Magnetic field components and magnitude. (c) Ion density. (d) Radial ion velocity component. (e) Tangential and normal ion velocity components. (f) Parallel, perpendicular and total ion temperatures. The location of a main magnetic switchback is marked in panel (c). The reconnection exhausts listed in Table~\ref{table:tab_1} are marked with green arrows (panels d and e).}
          \label{fig:fig_1}
\end{figure*}

We analyze observations from the Solar Orbiter mission on 16 July 2020 during a slow Alfvénic solar wind interval at a distance of about 0.64 AU (see \citet{DAMICIS_2021_AA}, this issue). This period was chosen because it contains substantial amounts of multiple proton populations, as explained next. We primarily make use of in situ data from the Solar Orbiter SWA \citep[Solar Wind Analyzer;][]{OWEN_2020} instrument suite that provides ion (PAS: Proton and Alpha Sensor) and electron (EAS: Electron Analyzer System) 3D velocity distribution functions with cadences up to 4 Hz and high angular and energy resolutions. We also use magnetic field data from the magnetometer \citep[MAG][]{HORBURY_2020} instrument with a cadence of $8$ Hz. 

\begin{figure*}
   \centering
   \includegraphics[width=1\textwidth]{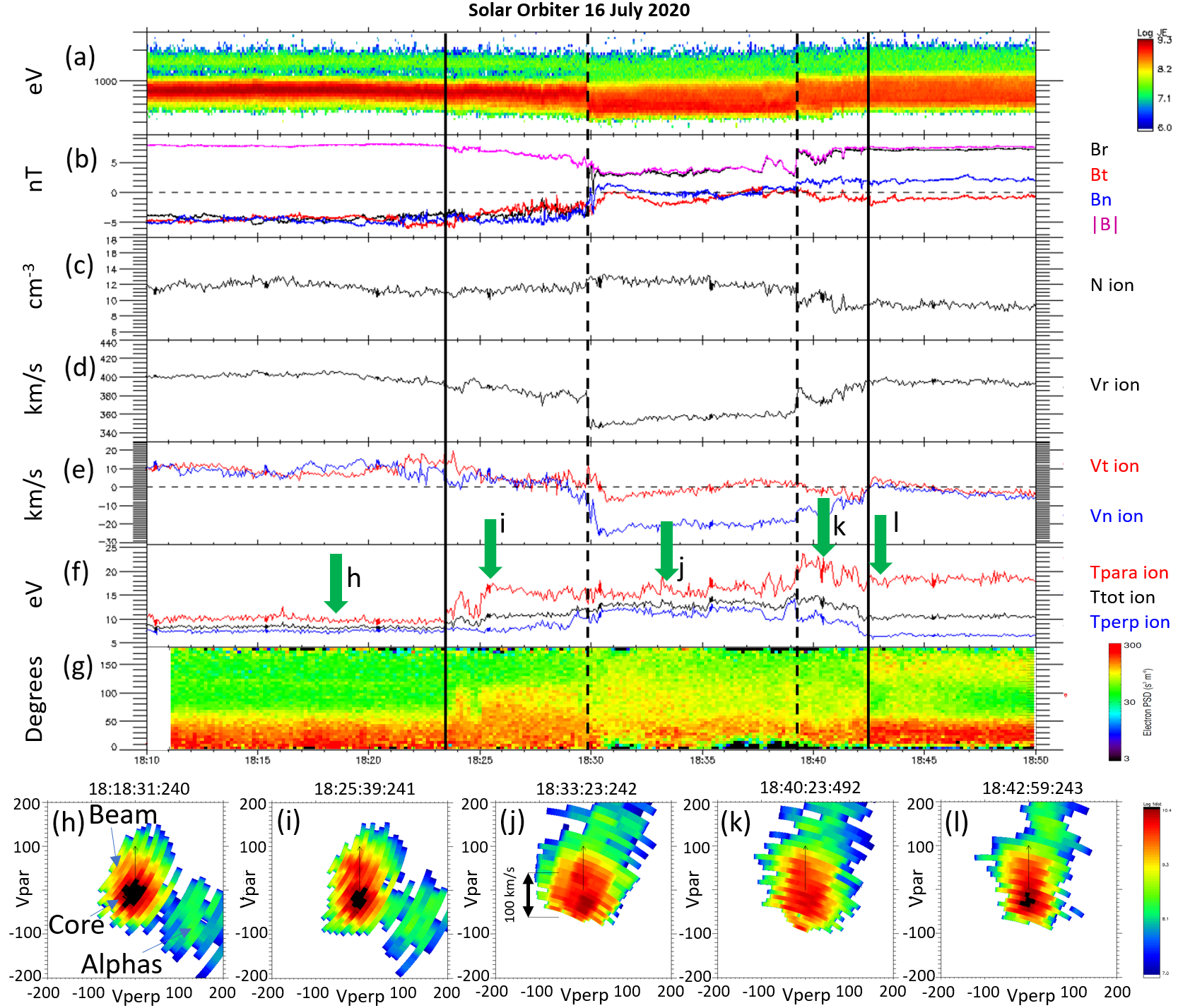}
   \caption{Solar Orbiter ion and magnetic field observations zoomed on the vicinity of the reconnection event around 18:35 UT on 16 July 2020 (Event 4 in Table~\ref{table:tab_1}). (a) Energy time spectrogram for all ions. (b) Magnetic field components and magnitude. (c) Ion density. (d) Radial ion velocity component. (e) Tangential and normal ion velocity components. (f) Parallel, perpendicular and total ion temperatures. (g) Electron pitch angle distribution above 75 eV. The reconnection exhaust (based on flows and current sheets) is comprised in between the vertical dashed lines. The extents of the external ion boundary layers are marked with solid vertical black lines. In panels (h) through (l), five cuts of the ion distribution functions in the Vpar-Vperp plane are shown. Thin arrows show the magnetic field direction, which is always toward the top in such Vpar-Vperp display. Vperp is defined along the convection direction and the distributions are displayed in the plasma bulk flow frame. Corresponding times are marked with green arrows in panel (f). These are characteristic ion distribution functions for each interval shown, as detailed in the text. In distribution (h), the locations of the proton core, proton beam, and alpha populations are highlighted for clarity. The phase space density is displayed with the same color scale for all distributions.}
              \label{fig:fig_2}
    \end{figure*}

Figure~\ref{fig:fig_1} displays the data for 16 July 2020. Panel (a) often shows a broad ion energy spectrum, in particular around the beginning and end of the interval, consistent with the larger parallel ion temperature observed in panel (f). This larger parallel temperature does not represent a mere bulk heating, but rather signals the presence of an ion beam on top of the core proton population, aligned with the magnetic field. During radial magnetic field periods (such as the end of this interval for instance), this translates into the beam being observed clearly in the energy spectrum of panel (a). The multiple proton populations are observed very clearly in the 3D ion distribution functions, as exemplified later in Figures~\ref{fig:fig_2}h and \ref{fig:fig_2}l for instance.

Although not central to the present study, we note the presence of several magnetic switchbacks \citep{Kasper_2019, Bale_2019} during this interval. We remind here that such large-scale Alfvénic structures have been observed in the past \citep{1971JGR....76.3534B} and that their steepening was proposed to explain the formation of sharp discontinuities with magnetic decreases at their edges (e.g., \citet{1999RvGeo..37..517T}; \citet{2002GeoRL..29.2233T}). The main switchback signature during the present interval is marked in panel (c) of Figure~\ref{fig:fig_1}. 

During most of the day, the magnetic field and velocity vectors are strongly anti-correlated, as expected for such a slow Alfvénic wind interval (cf. \citet{DAMICIS_2021_AA}). The periods of switchbacks, where the radial magnetic field component reverses, are accompanied by increases in the radial velocity component. These increases are not as strong as expected, however (from pure Alfvénicity; not shown). These may yet be called magnetic switchbacks since the magnetic field radial component does reverse and strahl electrons do not change direction (as discussed next with regards to Figure~\ref{fig:fig_2}). The question of the exact definition of switchbacks, however, is not the scope of the present paper.

A main feature stands out, however, upon exit of the main switchback, at 18:35:00 UT and marked by a green arrow in panel (d). A decrease in the magnetic field, associated with a strong decrease in radial velocity, marks the presence of a reconnection jet at the exit boundary of the switchback, confirming recent observations by \citet{FROMENT_2021} (see also \citet{FEDOROV_2021}, this issue). 

Figure~\ref{fig:fig_2} focusses on this reconnection event. It shows the same observations as in Figure~\ref{fig:fig_1}, here complemented by suprathermal electron pitch angle distributions for energies above 75 eV and several cuts of the ion distribution functions in key regions. We note that the electron data (panel g) demonstrate that the directionality of the strahl (the field-aligned population near $0^o$) does not change across the event as the radial magnetic field changes from sunward to anti-sunward, consistent with the interpretation of the event highlighted in Figure~\ref{fig:fig_1} as a large switchback.

The reconnection event shows a clear bifurcated current sheet structure embedding a jet with decreased radial speed. The spacecraft is thus located sunward of the X-line. The exhaust boundaries are marked with vertical dashed black lines, and lasts about 9:30 min. It shows an anti-correlation between ion velocity and magnetic field upon entry and a correlation upon exit (cf. for this event also \citet{DAMICIS_2021_AA}). Clear electron (broader pitch angle distributions) and ion boundary layers (signatures in ion flows, temperatures and distribution functions in particular) are also observed outside the exhaust boundaries, as reported in previous solar wind reconnection studies \citep{Huttunen_2008, LAVRAUD_2009}. The full extent of the exhaust and its boundary layers is $19$ min (cf. Table~\ref{table:tab_1}), as marked with solid vertical black lines. 

Using Minimum Variance Analysis \citep[MVA; e.g.][]{SONNERUP_1967} on this event, \citet{DAMICIS_2021_AA} (this issue) determined a normal to the reconnecting current sheet directed as $[-0.49, -0.18, 0.85]$ in RTN coordinates. Taking a mostly radial bulk solar wind speed of 400 km/s (outside the exhaust), an exhaust duration of 9:30 min, but with a spacecraft trajectory accounting for the normal orientation given above, one finds an exhaust width of $111720$ km. Assuming an opening angle of the reconnection exhaust in the solar wind following a typical X-line aspect ratio of $10$, this means the spacecraft is crossing the exhaust about $10^6$ km away from the X-line, though this type of calculation likely underestimates the distance. The total magnetic shear across the exhaust is about $125^o$. The normal magnetic field component is found to be close to zero but generally negative. The ratio of the magnetic field along the $M$ and $L$ components, or guide field, is on the order of $0.5$. \citet{DAMICIS_2021_AA} also performed the Walén test \citep{HUDSON_1970, PASCHMANN_1979} on this event, which showed very nicely Alfvénic and anti-correlated magnetic field and velocity components upon entry, and correlated upon exit.

We now focus on the details of the ion distribution functions in these various regions. Distribution (h) at the bottom of Figure~\ref{fig:fig_2} corresponds to the slow Alfvénic solar wind ahead and upstream of the exhaust. It shows a core proton distribution with the addition of a significant shoulder in the field-aligned direction. This type of distributions is often observed in the solar wind and is typically interpreted as the result of the presence of a lower density proton beam aligned with the magnetic field (cf. introduction). Note that as highlighted in distribution (h), the solar wind alpha particle population is also observed at higher velocities (i.e., energies). This is due to the fact that the PAS instrument measures particles as a function of energy per charge, so that alphas are measured at higher energies than protons. Note that the alpha particle population is also observed in Figure~\ref{fig:fig_2}a, for example as the green population around 1.5 keV at the beginning of the interval. In distribution (h) the proton beam is measured sunward of the proton core, which is consistent with the fact that the field lines are folded back towards the Sun at this time, as compared to distribution (l) outside the switchback, where the proton beam is flowing anti-sunward faster than the core distribution (also along the magnetic field). This is consistent with a switchback scenario where ions follow the kinks of the field lines, as already explained by \citet{NEUGEBAUER_2013}. Note that the ion population at higher velocities in the bottom-right quadrant of  distribution (h) corresponds to the solar wind alpha population. It is also seen in the top-right quadrants of distributions (j), (k) and (l). We do not analyse this population further in the present study.

Upon entry of the spacecraft into the outer boundary layer (ion separatrix), at the first solid vertical black line, the ion distribution function (i) significantly changes compared to previous times (h). A much stronger and faster (relative drift) proton beam, caused by reconnection, is observed flowing away from the exhaust (along the magnetic field direction). As the spacecraft enters the exhaust, through the first vertical dashed line, the distribution function drastically changes again. Distribution function (j) indeed shows two proton populations of similar intensities (rather than a beam and stronger core), drifting relative to each other along the magnetic field. Such well resolved beams in a solar wind reconnection exhaust had been reported so far only in the seminal paper of \citet{GOSLING_2005}, although parallel ion heating likely due to interpenetrating beams produced by reconnection has been observed \citep[e.g.][]{PHAN_2006, PHAN_2021, HE_2018, EASTWOOD_2018}. 

As explained by \citet{GOSLING_2005}, within the exhaust the separation in velocity space between the two populations is expected to be the sum of the two upstream Alfvén speed ($V_{A1} + V_{A2}$) because this is the speed of the kink through which each of the beam is penetrating to form the exhaust. For the event of Figure~\ref{fig:fig_2}, the Alfvén speed is on the order of $V_{A1} \sim 49$~km~s$^{-1}$ before entry and $V_{A2} \sim 54$~km~s$^{-1}$ upon exit (outside the thick vertical black lines), yielding an expected separation between the populations along the magnetic field of a bit more than a $100$~km~s$^{-1}$. This is on the order of the separation observed in Figure~\ref{fig:fig_2}j, albeit somewhat larger. As already noted in \citet{GOSLING_2005}, the lower separation is consistent with the electromagnetic ion beam instability which may limit the relative speed to $\sim 1.5 V_A$ \citep{GOLDSTEIN_2000}, roughly as observed.

After the spacecraft exits into the second boundary layer (second vertical dashed line), ion measurement display yet more complex structures with increased parallel and perpendicular temperatures (panel 2f). The distribution function of Figure~\ref{fig:fig_2}k, representative of that region, shows the presence of possibly more than 2 populations, including a small beam possibly propagating anti-parallel to the magnetic field direction (towards the bottom). This is consistent with protons exiting the exhaust and thus propagating anti-parallel to the magnetic field on this side of the reconnection exhaust. 

 \begin{figure*}
   \centering
   \includegraphics[width=6.5in]{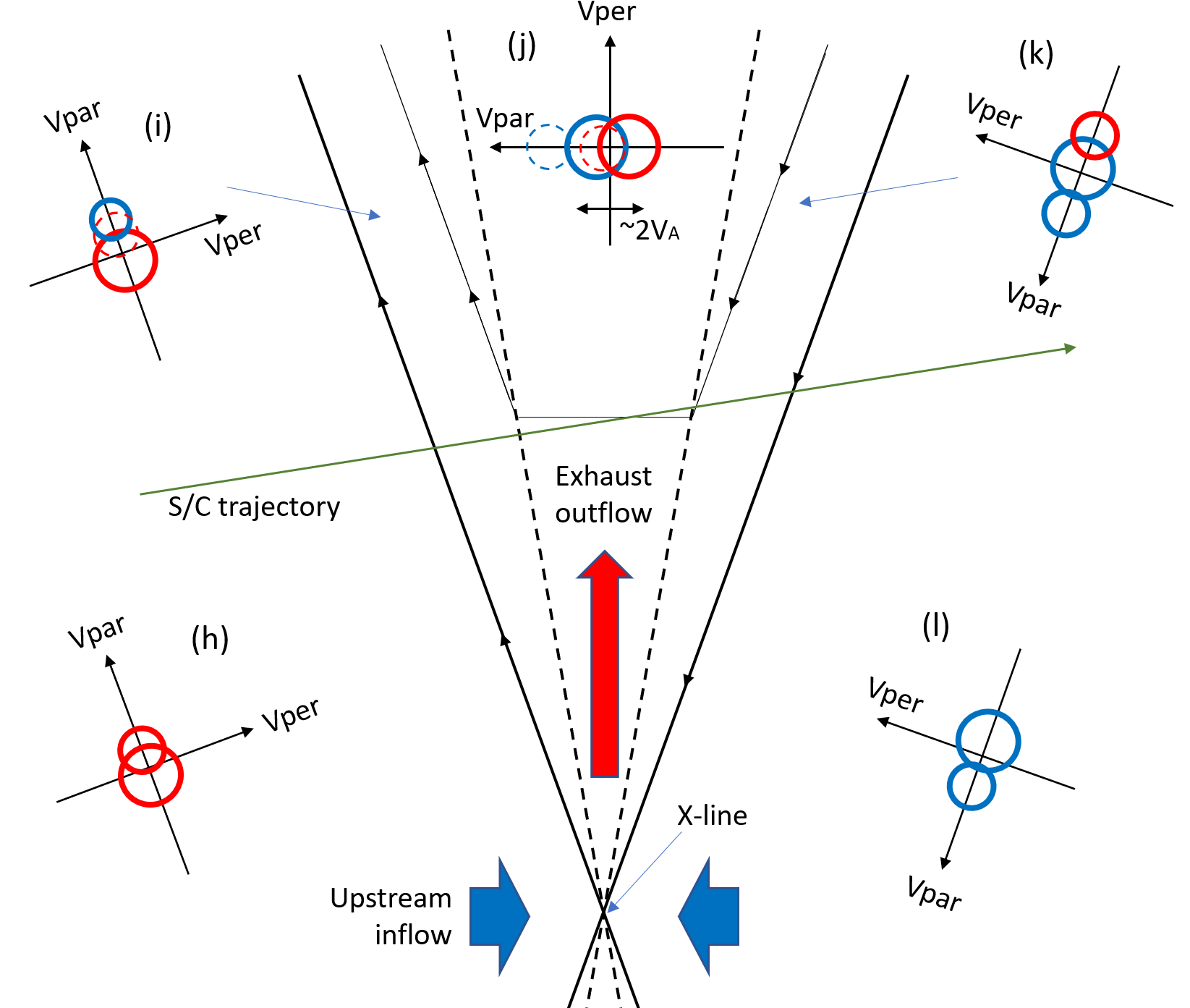}
   \caption{Schematic description of the distribution functions and magnetic reconnection geometry observed in the vicinity of the event analyzed in Figure~\ref{fig:fig_2}. The schematic distributions are meant to be in the plasma frame, as for the distribution displayed in Figure~\ref{fig:fig_2}. Initial upstream proton populations are colored red and blue. Circles are used to represent the populations in phase space, and their diameter is meant to represent the intensity of the population. Their mixing in the various regions is illustrated through the use of the same color coding to identify the origin of each population. The magnetic field lines are shown with solid black lines (thick ones show the separatrices) and the exhaust boundaries (current sheets) are shown with dashed lines, thus using the same coding as for the vertical lines of Figure~\ref{fig:fig_2}. It should be noted that the pre-existence of beams in the upstream regions is not required for reconnection to occur, and that the exact details of the mixing of the various populations given in this figure is simplified as it does not account for kinetic behaviors that would require dedicated analysis \citep[e.g.][]{COWLEY_1982}. See text for further details.}
              \label{fig:fig_3}
    \end{figure*}

Although detailed kinetic modeling is left for future work, Figure~\ref{fig:fig_3} provides a schematic description of the initial proton populations and their observed mixing in the various regions associated with the reconnection event of Figure~\ref{fig:fig_2}. The proton distribution functions on each side of the reconnecting current sheet are grossly sketched using circles whose size represent the beam intensity. These are colored red and blue, respectively for the left and right upstream regions. The upstream populations both show the presence of a core and a beam, with different relative drifts and intensities (cf. Figures \ref{fig:fig_2}h and \ref{fig:fig_2}l and their counterpart in Figure~\ref{fig:fig_3}). As discussed in the next section, the presence of pre-existing beams in upstream populations is of interest for reconnection dynamics, but is of course not needed a priori for reconnection to occur.

As the spacecraft crosses into the first boundary layer a field-aligned population of stronger intensity and larger relative drift appears (Figures \ref{fig:fig_2}i and \ref{fig:fig_3}i). It is interpreted as a population produced by reconnection and likely originates from the other side of the exhaust. This population is thus colored in blue in distribution (i) of Figure~\ref{fig:fig_3}, and it likely overlaps with the pre-existing proton beam population on this side of the exhaust, which is represented by a red circle but is dashed since it is likely hidden by the population produced by reconnection, and not observable. In the exhaust itself, Figure~\ref{fig:fig_2}j is made of two main populations of similar intensities, which are interpreted as coming from the mixing of the two main upstream proton core populations. In Figure~\ref{fig:fig_2}j, a longer tail is observed towards higher field-aligned velocities, which is consistent with the remnants of the proton beam from the right-hand side (small blue circle) population while the proton beam from the left-hand side is likely hidden by the other populations and is thus represented by a dashed red circle. The distribution function of Figure~\ref{fig:fig_2}k in the boundary layer observed upon exit is more complex, with overall larger temperatures, both parallel and perpendicular to the magnetic field. This is interpreted as the superposition of proton core and beam populations from each side as described in distribution k of Figure~\ref{fig:fig_3}. A proton beam coming from the other side of the exhaust is expected to propagate this time anti-parallel to the magnetic field (small blue circle towards negative parallel velocities). This later population is somewhat less well resolved. It likely is due to the effect of wave particle interactions that are likely at work for such structured distributions. Higher angular and energy resolutions would be needed to ascertain the presence of all the populations sketched in Figure~\ref{fig:fig_3}k.

In addition to this reconnection event, we found six other clear reconnection cases during that day. They are listed in Table~\ref{table:tab_1}. Only the event of Figure~\ref{fig:fig_2} occurred at the boundary of a switchback. When taking both the exhausts and boundary layers (when identified) of all six events the total duration amounts to 28:20 min, or about 2\% of the measurements of this 1-day interval. This is a minimum estimate because current sheets at 03:15:20, 05:47:50, 06:06:00, 14:28:15, 22:02:00, and 23:21:10 UT are ambiguous, with possible reconnection signatures (decreased magnetic field, bifurcated current sheet, ion jets or multiple component proton distributions). There are also two clean bifurcated current sheets at 07:01:45 UT and 07:04:55 UT, which are likely reconnecting, but occurring during a data gap in ion measurements. Finally, there might be other current sheets of interest, but too small or too low shear to be properly identified.

\begin{table}
\caption{List of confirmed reconnection events on 16 July 2020, with the duration of the interval that includes the exhaust and its boundary layers, and presence or absence of associated multiple populations with Tpara > Tperp in the distribution functions.}              % title of Table
\label{table:tab_1}      % is used to refer this table in the text
\centering                                      % used for centering table
\begin{tabular}{|c|c|c|c|}          % centered columns (2 columns)
\hline                      % inserts double horizontal lines
Event & Time (hh:mm:ss) & Duration (mm:ss) & Multiple pop. \\    % table heading
\hline                                   % inserts single horizontal line
1 &	06:56:50 &  01:30 &	Yes \\ 
2 &	12:36:02 &	06:00 &	Likely \\
3 &	14:56:30 &	00:30 &	Yes \\
4 &	18:35:00 &	19:00 &	Yes \\
5 &	21:56:50 &	00:20 &	Yes \\
6 &	23:01:30 &	01:00 &	Yes \\
 \hline
\end{tabular}
\end{table}

%%%%%%%%%%%%%%%%%%%%%%%%%%%%%%%%%%%%%%%%%%%%%%%%%%%%%%%%%%%%%%%%%%%%%%%%%%%%%%
\section{Discussion} \label{sec:3_dis}
%%%%%%%%%%%%%%%%%%%%%%%%%%%%%%%%%%%%%%%%%%%%%%%%%%%%%%%%%%%%%%%%%%%%%%%%%%%%%%

The reconnection event in Figure~\ref{fig:fig_2} constitutes a textbook example of multiple proton populations and beam production by magnetic reconnection locally in the solar wind. Yet, this event has special boundary conditions, with pre-existing multiple populations on either side of the exhaust in the upstream regions. Had the event been observed in solar wind conditions with no such beams, only the reconnection exhaust and its boundary layers would have featured multiple component populations and beams, as the sole result of reconnection. Indeed, the pre-existence of such beams in upstream conditions is not needed for reconnection to proceed. However, we shall recall here that recent work \citep{ALTERMAN_2019} suggests that about 70\% of solar wind proton measurements display discernable multiple populations, so that such conditions are likely preponderant in the solar wind.

The fact that magnetic reconnection may create multiple proton populations is known from some time, from both observations and simulations \citep[e.g.][]{LOTTERMOZER_1998, SHAY_1998, HOSHINO_1998, PHAN_2007, DRAKE_2009, AUNAI_2011, EASTWOOD_2015, INNOCENTI_2017}. For solar wind configurations, complex reconnection exhaust boundaries have been observed and modeled \citep[e.g.][]{Huttunen_2008, LAVRAUD_2009, INNOCENTI_2017, FENG_2017, ENZL_2017, HE_2018}. 

What the observations of Figure~\ref{fig:fig_2} show is that during a large portion of the time, multiple proton populations already exist in the solar wind and can constitute the boundary conditions of magnetic reconnection. The high-resolution Solar Orbiter ion measurements nicely highlight that very complex distributions may arise from such conditions in the exhausts and its boundary layers. The structure of the distribution functions in each region associated to the reconnection geometry can be explained by the mixing and superposition of the upstream populations, as described in Figure~\ref{fig:fig_3}. Future, more detailed theoretical and simulation works should aim to confirm this scenario, and further constrain the kinetic behavior of protons associated with reconnection in the solar wind in the presence of pre-existing multiple populations. In particular, determining exactly which parts of the proton populations are able to cross each given boundary, and mix or escape, require a kinetic treatment of the proton properties \citep[e.g.][]{COWLEY_1982}.

Solar wind reconnection exhausts are rather ubiquitous in the solar wind, even if not omnipresent. They are known to have significant extents both along the X-line and along the exhaust, and to be operating at times for hours \citep{PHAN_2006, GOSLING_2007_GRL, LAVRAUD_2009}. The jet orientation and proximity to the X-line are two other random factors that influence the time spent by the spacecraft inside the exhaust and its boundary layers (note the large differences in duration in Table~\ref{table:tab_1}). Also, it remains unknown whether beams produced by reconnection may be observed at large distances in the solar wind without local signatures of reconnection (such as local bifurcated current sheets or ion jets). In addition, flux ropes, folded magnetic fields, and switchbacks are all structures potentially related to magnetic reconnection in the corona or solar wind. Multiple proton populations in the vicinity of such structures are thus also possibly the result of magnetic reconnection, suggesting the role of this process may not be limited to exhaust regions (bifurcated current sheets) and their separatrices, but could produce broader regions of multiple ion populations if produced closer to the Sun. All these facts conspire to suggest that reconnection may be a non-negligible process in the production of multiple proton populations in the solar wind, despite the apparent modest 2\% contribution found in the present study. Future works are needed to determine this possibility more quantitatively. 

From multi-spacecraft observations, \citet{GOSLING_2007_APJL} found that current sheets, and thus also reconnection X-lines, in the turbulent, high-speed wind are considerably more localized than in the low speed wind or in interplanetary coronal mass ejections. It must be noted that proton beams have been observed in all types of solar wind. The solar wind interval studied here, on 16 July 2020, corresponds to a specific type of slow Alfvénic solar wind associated to a pseudostreamer (cf. \citet{DAMICIS_2021_AA}, this issue). Such type of wind resembles fast solar wind in various respects \citep[e.g.][]{DAMICIS_2019, DAMICIS_2021, STANSBY_2020}. By contrast, the regular slow solar wind is known to be formed of various structures, current sheets and flux ropes likely resulting from reconnection closer to the Sun \citep[e.g.][]{VIALL_2008, VIALL_2015, KEPKO_2016, SANCHEZ_DIAZ_2019, LAVRAUD_2020}. The slow solar wind has been shown to contain larger amounts of proton beams \citep[e.g.][]{ALTERMAN_2019}. The proportion of multiple proton populations produced locally by magnetic reconnection might thus be larger in the slow solar wind. 

Early works by \citet{GOSLING_2006a, GOSLING_2006b} have shown that magnetic reconnection inside $1$ AU (down to $0.3$ AU with Helios) is somewhat less frequent, while observations beyond 1 AU shows solar wind reconnection at large heliocentric distances occurs about as frequently and with similar characteristics as at $1$ AU (apart from some properties related to the higher $\beta$). That the observation of magnetic reconnection exhausts in the solar wind closer to the Sun is less frequent remains to be quantified more thoroughly on the basis of Parker Solar Probe (PSP) data analysis \citep[e.g.][]{PHAN_2020}, albeit we already know it is prevalent at the Heliospheric Current Sheet close to the Sun \citep{SZABO_2020, LAVRAUD_2020, PHAN_2021}. Therefore, the proton beams regularly observed close to the Sun with PSP are most likely unrelated to reconnection occurring locally in the solar wind \citep[e.g.][]{VERNIERO_2020}, though a reconnection-related process down in the low corona remains one of the proposed mechanisms \citep[e.g.][]{1996A&A...316..355F}. It was noted earlier that proton beams in the fast solar wind are generally less resolved as the heliocentric radial distance increases \citep{MARSCH_1982}. These arguments all suggest that the proportion of multiple proton populations produced by magnetic reconnection locally in the solar wind should increase with distance from the Sun. 

Although not the subject of the present paper, we also finally note that as distance from the Sun increases, proton beams produced at interplanetary and planetary shocks are also expected to populate the solar wind.

%%%%%%%%%%%%%%%%%%%%%%%%%%%%%%%%%%%%%%%%%%%%%%%%%%%%%%%%%%%%%%%%%%%%%%%%%%%%%%
\section{Conclusions} \label{sec:4_conclusions}
%%%%%%%%%%%%%%%%%%%%%%%%%%%%%%%%%%%%%%%%%%%%%%%%%%%%%%%%%%%%%%%%%%%%%%%%%%%%%%

We have analyzed recent high-resolution ion, electron and magnetic field observations from the Solar Orbiter mission during an interesting interval of slow Alfvénic solar wind on 16 July 2020. We focused on the presence of multiple proton populations and beams. We found that (1) high-resolution Solar Orbiter observations confirm magnetic reconnection as a ubiquitous process in the solar wind, (2) that it produces interpenetrating proton populations of equivalent intensities in the exhaust, as well as (3) lower density beams outside the exhaust along ion separatrix layers, (4) and that it therefore constitutes a non-negligible mechanism to produce multiple proton populations and beams in the solar wind. We find that 2\% of the multiple populations observed during the 1-day interval under study are related to magnetic reconnection. We discussed reasons to believe that this proportion may be underestimated, and that it should depend on the type of solar wind and distance from the Sun. This study suggests that magnetic reconnection is not a dominant process for producing such proton distributions in the solar wind. The found proportion may be viewed as very little, or much, given that this process has not been considered at all so far. Future studies shall assess its role more quantitatively, and Solar Orbiter is particularly well geared for that purpose. 

Three interesting side conclusions of the present study are the confirmation that: (1) magnetic reconnection in the solar wind should often involve upstream conditions that are already made of multiple particle populations, (2) the multiple proton beam structures follow the magnetic field geometry as it folds back on itself in switchbacks, as first explained by \citet{NEUGEBAUER_2013}, and (3) magnetic reconnection does occur at the boundaries of at least some switchbacks \citep{FROMENT_2021}.

%%%%%%%%%%%%%%%%%%%%%%%%%%%%%%%%%%%%%%%%%%%%%%%%%%%%%%%%%%%%%%%%%%%%%%%%%%%%%%
%%%%%%%%%%%%%%%%%%%%%%%%%%%%%%%%%%%%%%%%%%%%%%%%%%%%%%%%%%%%%%%%%%%%%%%%%%%%%%

\begin{acknowledgements}
    Solar Orbiter is a space mission of international collaboration between ESA and NASA, operated by ESA. We thank all the ESA, NASA and instrument team members who made Solar Orbiter a success. We are grateful to the International Space Science Institute (ISSI) for its support of the team "Unravelling solar wind microphysics in the inner heliosphere" dedicated in part to the analysis of Solar Orbiter data. Work at LAB and IRAP was supported by CNES and CNRS. Solar Orbiter magnetometer operations are funded by the UK Space Agency (grant ST/T001062/1). T.S.H. and J.P.E. are supported by STFC grant ST/S000364/1. D.V. is supported by STFC Ernest Rutherford Fellowship ST/P003826/1 and STFC Consolidated Grant ST/S000240/1. L.P. was supported by the Czech Science Foundation, grant no. 19-18993S.
\end{acknowledgements}

%%%%%%%%%%%%%%%%%%%%%%%%%%%%%%%%%%%%%%%%%%%%%%%%%%%%%%%%%%%%%%%%%%%%%%%%%%%%%%
\bibliographystyle{aa}
\bibliography{SOURCES}

\end{document}